# Super-Dispersive Demultiplexer Design Using Positive-Negative Refraction Boundary and Hetero-Photonic Crystals

Saeed Pahlavan
Electrical and Computer Engineering Faculty
Tarbiat Modares University
Tehran, Iran
s_pahlavan@modares.ac.ir

*Abstract*—In this paper, a systematic approach is employed to design a photonic crystal with a boundary between positive and negative refraction to boost the refractive properties of the crystal. Mathematical techniques are employed to turn the process of EFC engineering into a nearly mechanical process free of trial-and-error steps. The designed demultiplexer's operational characteristics are then further improved by adding a second photonic crystal with an oblique boundary. The refracted beams in the first crystal impinge on an oblique interface of the second photonic crystal to experience a change in Bloch wavenumbers and even greater refraction angles so that a novel super-dispersive two-step optical demultiplexer is made. A beam divergence of 161 degrees is obtained for an input spectrum of λ= [1474nm, 1550nm].

## I. Introduction

The key to designing a high-performance optical demultiplexer is to improve the refractive properties of the demultiplexing medium. From the early studies of photonic crystals, their refractive properties have been in the spotlight [1] and their suitable refractive properties through the super-prism effect have been highlighted [2] [3]. Inspired by the potential of photonic crystals for demultiplexing, several groups tried to shed light on the problem of light refraction in photonic crystals [4] [5] [6] and the employment of super-prism effect in their structures [7] [8] [9] [10] in order to produce high-performance demultiplexers which are a key component of communication systems at the optical wavelengths as well as other regions of the electromagnetic spectrum [11]. These studies led to the advent of the comprehensive theory of light refraction in photonic crystals and the role of equi-frequency contours (EFC) in determining the trajectory of incident light inside the crystal [12] [13]. Design of photonic crystals for the required EFC is essentially a procedure based on trial-and-error, but a mathematical design strategy has been proposed to turn this procedure into a step-by-step nearly mechanical process [14]. In this paper, by employing this strategy, a boundary between positive and negative refraction for the incident spectrum in the photonic crystal is created to enhance wavelength separation in the demultiplexer. The separated wavelengths then impinge on a second crystal with an oblique boundary to change their Bloch wavenumbers and increase the separation angles even further according to the technique described in [15].

The paper is organized as follows: The next section explains the theory of light refraction in photonic crystals and the mathematical tools and steps needed for EFC engineering. Designs of demultiplexers exhibiting positive-negative refraction boundaries are described in the third section. The fourth section includes the designs of two-step demultiplexers enjoying a positive-negative refraction boundary in the first step and a change of excited Bloch wavenumbers due to oblique interfaces of photonic crystals in the second step. The last section concludes the paper.

## II. Theory

Light in photonic crystals can be directed to propagate in such a way as if the medium has a negative refractive index. This necessitates a thorough investigation of the rules of light refraction in photonic crystals if we are to use this phenomenon in our designs. This problem was solved for ordinary homogeneous and isotropic media a long time ago by Snell's rule, but due to the periodic modulation of electric permittivity in photonic crystals, these structures don't obey Snell's rule.

Bloch-Floquet theory is the dominant theory for the explanation of light propagation problems inside photonic crystals, but it only determines the allowed propagation frequencies but not the light trajectory. Other types of periodic structures at other regions of electromagnetic spectrum can also have various applications and there are other approaches for their analysis [16]. Several theories were proposed by different groups with the goal of light trajectory prediction in photonic crystals but all of them proved to be insufficient [1] [6] [17] [18] [19] [20] [21]. Eventually, a paper by Notomi shed light on the subject by providing a theory that was in good agreement with experiments [12]. Refracted light trajectory in photonic crystals is determined with the help of the EFCs, a mathematical tool that is the heart of the theory of [12] and is the locus of all of the wavevectors in the reciprocal lattice which share a certain frequency of propagation as their eigenvalue.

To determine the propagation angles of the excited waves in a photonic crystal, the EFC of the incident light material should be drawn on top of the EFC of the photonic crystal. Figure 1 shows the EFC of an isotropic and homogeneous medium (top) and the schematic EFC of a 2D photonic crystal (bottom). For a homogeneous and isotropic medium, we have:

$$\omega = ck/n = c\sqrt{k_x^2 + k_y^2}/n \qquad (1)$$

where $\omega$ and $c$ are the angular frequency and the speed of light, respectively; $k$ is the wavenumber and $n$ is the refractive index of the medium. This equation shows that for a homogeneous and isotropic medium, the points in $k_x$-$k_y$ space that correspond to a certain frequency form a circle, which is the reason why the top EFC is circular in Fig. 1. The vertical line drawn from the $k_x$ point of the incident wave determines the value of $k_x$ in the photonic crystal because the tangential components of the electric field are the same in the two media to satisfy electromagnetic boundary conditions. Thus, The Bloch wavenumber of the excited wave in the photonic crystal is known. The propagation trajectory of the excited wave in the photonic crystal is in the direction of the group velocity vector which is the gradient of the angular frequency with respect to the wavevector $v=\text{grad}_\kappa\omega$ which is a line perpendicular to the EFC.

The group velocity direction is obviously the direction of the gradient: if the EFCs are growing by frequency, the perpendicular line should be drawn outward, and if the contours are shrinking by increasing frequency, the line should be drawn inward. For the case depicted in Fig. 1, if the EFC of the photonic crystal has an inward gradient, the incident light propagates in a way as if the crystal has a negative refractive index. Therefore, in order to have a photonic crystal with negative refraction, we should design a crystal whose EFCs at the wavelength range of interest show an inward gradient.

The key idea towards EFC engineering is to use a specific kind of crystal that somehow eases the design procedure. One such crystal is a staircase profile for which closed-form relations exist between frequency and crystal parameters [22]. We consider a two-dimensional rectangular stair-case (RSC) crystal whose permittivity function obeys [22]:

$\varepsilon(x,y)= A(x)+B(y)$ (2)

where $A$ and $B$ are stair-case profiles:

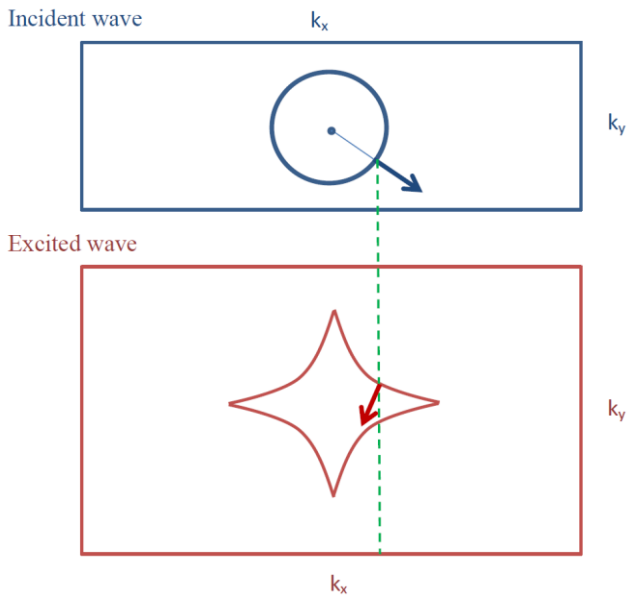

Fig. 1. EFCs of an isotropic and homogeneous medium (top) and schematic EFC of a photonic crystal (bottom). The $k$ vector in the crystal is determined by the continuity of the tangential components across the boundary. The light propagation direction inside the photonic crystal is oriented towards the gradient of the EFC.

$A(x)=\varepsilon_a+[u(x+t/2)-u(x-t/2)](\varepsilon_b-\varepsilon_a)$
$|x|\le X/2 \qquad A(x)=A(x+mX)$ (3-a)

and

$B(y)=\varepsilon_c+[u(y+s/2)-u(y-s/2)](\varepsilon_d-\varepsilon_c)$
$|y|\le Y/2 \qquad B(y)=B(y+nY)$ (3-b)

$u$ is the unit step function and $X$ and $Y$ are the periods of the functions along the $x$ and $y$ directions. $m$ and $n$ are integers and $t$ and $s$ are positive constants for which we have: $t<X$ and $s<Y$. A schematic of such a crystal is shown in Fig. 2.

Two-dimensional wave equation for TE waves reads:

$(\frac{\partial^2}{\partial x^2}+\frac{\partial^2}{\partial y^2})E(x,y)+\frac{\omega^2}{c^2}\varepsilon(x,y)E(x,y)=0$ (4)

where $E(x,y)$ is the electric field amplitude and $\omega$ and $c$ are the frequency and speed of light. Using the permittivity function of equation (2) and assuming $E(x,y) = \Lambda(x)\Psi(y)$ and doing some algebraic manipulations we reach:

$\frac{\partial^2}{\partial x^2}\Lambda(x)+\frac{\omega^2}{c^2}[A(x)+\beta]\Lambda(x)=0$ (5-a)

$\frac{\partial^2}{\partial y^2}\Psi(y)+\frac{\omega^2}{c^2}[B(y)-\beta]\Psi(y)=0$ (5-b)

where $\beta$ is a separation constant. $A(x)+\beta$ and $B(y)-\beta$ are periodic functions by periodicities of $X$ and $Y$. Therefore, equations 5-a and 5-b have Bloch solutions of the form $\Lambda_\kappa(x)=\lambda_\kappa(x)\exp(-j\kappa x)$ and $\Psi_\eta(y)=\psi_\eta(y)\exp(-j\eta y)$, where $\kappa$ and $\eta$ are Bloch wavenumbers. After lengthy manipulations we obtain:

$\cos(\kappa X)=\frac{q_{11}}{2}\exp\{-j\omega\frac{X}{c}[A(-\frac{X}{2})+\beta]^{1/2}\}+$

$\frac{q_{22}}{2}\exp\{+j\omega\frac{X}{c}[A(-\frac{X}{2})+\beta]^{1/2}\}$ (6-a)

$\cos(\eta Y)=\frac{p_{11}}{2}\exp\{-j\omega\frac{Y}{c}[B(-\frac{Y}{2})-\beta]^{1/2}\}+$

$\frac{p_{22}}{2}\exp\{+j\omega\frac{Y}{c}[B(-\frac{Y}{2})-\beta]^{1/2}\}$ (6-b)

$q_{ii}$ and $p_{ii}$ are the diagonal elements of the transfer matrix in the $x$ and $y$ directions which are related to the electrical permittivity function. Inserting the relations for the diagonal elements of the transfer matrix into equations 6-a and 6-b we have:

$\cos(\kappa X)=\cos(\frac{\omega}{c}\sqrt{\varepsilon_b+\beta}t)\cos[\frac{\omega}{c}\sqrt{\varepsilon_a+\beta}(X-t)]-$

$\frac{\varepsilon_a+\varepsilon_b+2\beta}{2\sqrt{\varepsilon_a+\beta}\sqrt{\varepsilon_b+\beta}}\sin(\frac{\omega}{c}\sqrt{\varepsilon_b+\beta}t)\sin[\frac{\omega}{c}\sqrt{\varepsilon_a+\beta}(X-t)]$

(7-a)

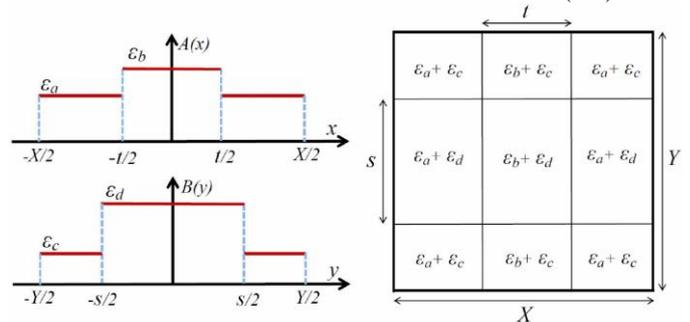

Fig. 2. Unit cell of the RSC profile used in the design

$$\cos(\eta Y) = \cos(\frac{\omega}{c}\sqrt{\varepsilon_d + \beta}s)\cos[\frac{\omega}{c}\sqrt{\varepsilon_c + \beta}(Y-s)] -$$
$$\frac{\varepsilon_c + \varepsilon_d + 2\beta}{2\sqrt{\varepsilon_c + \beta}\sqrt{\varepsilon_d + \beta}}\sin(\frac{\omega}{c}\sqrt{\varepsilon_d + \beta}s)\sin[\frac{\omega}{c}\sqrt{\varepsilon_c + \beta}(Y-s)]$$
(7-b)

We can set $X$ and $Y$ to unity for the sake of normalization and crystal design can be done by calculation of the degrees of freedom in these equations: $\varepsilon_a, \varepsilon_b, \varepsilon_c, \varepsilon_d, t, s$. Thus, we can write a set of 6 simultaneous nonlinear equations to solve for the six unknown variables:

1:
$$\cos(\kappa_1 X) = \cos(\frac{\omega}{c}\sqrt{\varepsilon_b + \beta}t)\cos[\frac{\omega}{c}\sqrt{\varepsilon_a + \beta}(X-t)] -$$
$$\frac{\varepsilon_a + \varepsilon_b + 2\beta}{2\sqrt{\varepsilon_a + \beta}\sqrt{\varepsilon_b + \beta}}\sin(\frac{\omega}{c}\sqrt{\varepsilon_b + \beta}t)\sin[\frac{\omega}{c}\sqrt{\varepsilon_a + \beta}(X-t)]$$

2:
$$\cos(\eta_1 Y) = \cos(\frac{\omega}{c}\sqrt{\varepsilon_d + \beta}s)\cos[\frac{\omega}{c}\sqrt{\varepsilon_c + \beta}(Y-s)] -$$
$$\frac{\varepsilon_c + \varepsilon_d + 2\beta}{2\sqrt{\varepsilon_c + \beta}\sqrt{\varepsilon_d + \beta}}\sin(\frac{\omega}{c}\sqrt{\varepsilon_d + \beta}s)\sin[\frac{\omega}{c}\sqrt{\varepsilon_c + \beta}(Y-s)]$$

3:
$$\cos(\kappa_2 X) = \cos(\frac{\omega}{c}\sqrt{\varepsilon_b + \beta}t)\cos[\frac{\omega}{c}\sqrt{\varepsilon_a + \beta}(X-t)] -$$
$$\frac{\varepsilon_a + \varepsilon_b + 2\beta}{2\sqrt{\varepsilon_a + \beta}\sqrt{\varepsilon_b + \beta}}\sin(\frac{\omega}{c}\sqrt{\varepsilon_b + \beta}t)\sin[\frac{\omega}{c}\sqrt{\varepsilon_a + \beta}(X-t)]$$

4:
$$\cos(\eta_2 Y) = \cos(\frac{\omega}{c}\sqrt{\varepsilon_d + \beta}s)\cos[\frac{\omega}{c}\sqrt{\varepsilon_c + \beta}(Y-s)] -$$
$$\frac{\varepsilon_c + \varepsilon_d + 2\beta}{2\sqrt{\varepsilon_c + \beta}\sqrt{\varepsilon_d + \beta}}\sin(\frac{\omega}{c}\sqrt{\varepsilon_d + \beta}s)\sin[\frac{\omega}{c}\sqrt{\varepsilon_c + \beta}(Y-s)]$$

5:
$$\cos(\kappa_3 X) = \cos(\frac{\omega}{c}\sqrt{\varepsilon_b + \beta}t)\cos[\frac{\omega}{c}\sqrt{\varepsilon_a + \beta}(X-t)] -$$
$$\frac{\varepsilon_a + \varepsilon_b + 2\beta}{2\sqrt{\varepsilon_a + \beta}\sqrt{\varepsilon_b + \beta}}\sin(\frac{\omega}{c}\sqrt{\varepsilon_b + \beta}t)\sin[\frac{\omega}{c}\sqrt{\varepsilon_a + \beta}(X-t)]$$

6:
$$\cos(\eta_3 Y) = \cos(\frac{\omega}{c}\sqrt{\varepsilon_d + \beta}s)\cos[\frac{\omega}{c}\sqrt{\varepsilon_c + \beta}(Y-s)] -$$
$$\frac{\varepsilon_c + \varepsilon_d + 2\beta}{2\sqrt{\varepsilon_c + \beta}\sqrt{\varepsilon_d + \beta}}\sin(\frac{\omega}{c}\sqrt{\varepsilon_d + \beta}s)\sin[\frac{\omega}{c}\sqrt{\varepsilon_c + \beta}(Y-s)]$$
(8)

$(\kappa_i, \eta_i)$, $i$=1,2,3 are points in the reciprocal lattice that we want to include in the EFCs. EFC engineering is done by a smart selection of these points to shape the EFCs for our desired purpose. We have employed numerical methods such as the Trust-Region-Dogleg method to solve the set of equations in (8) [23]. Note that due to the symmetries of the unit cell, the EFC profile can be shaped by shaping only a quarter of the $\eta$-$\kappa$ plane [24]. For example, for designing a circular EFC, it is adequate to design a 90° arc.

### III. DESIGN

Designing a crystal with a boundary between positive and negative refraction for the incident spectrum is a promising technique for a good demultiplexer design. As is schematically shown in Fig. 3, the change of refractive index is usually high in the boundary between positive and negative refraction and this produces a highly refractive material for the wavelength range of interest.

To design a crystal with a positive-negative refraction boundary, we choose our three points as follows:

$(\kappa_1, \eta_1) = (0, 0.1)$      for $\omega$=0.78
$(\kappa_2, \eta_2) = (0, 0.2)$      for $\omega$=0.74
$(\kappa_3, \eta_3) = (0, 0.3)$      for $\omega$=0.76      (9)

These points are selected in a way to guarantee the boundary between positive and negative refraction: Moving from point 1 to 2, the EFCs are growing with decreasing frequency, i.e., we have negative refraction, while moving from point 2 to 3, the EFCs are growing with increasing frequency and we have positive refraction. Thus, there would be a boundary between positive and negative refraction.

Solving for the crystal parameters using the Trust-Region-Dogleg method, we have:
$\varepsilon_a$=0.72, $\varepsilon_b$=1.82, $\varepsilon_c$=0.8, $\varepsilon_d$=1.97, s=0.56, t=0.62      (10)

EFCs of the designed RSC crystal are drawn with the help of the plane-wave expansion method and are shown in Fig. 4 [25]. As is clear in the figure, the selected points are successfully included in the EFCs.

The designed RSC crystal can be converted to standard pillar or air-hole structures which are easier to fabricate, with the help of mathematical techniques described in [22] and [14]. As described in [22] and [14], the standard crystal parameters can be calculated in a way so that its first few Fourier coefficients are approximately equal to those of the designed RSC crystal. As a result of satisfying this condition, the two crystals show nearly identical EFCs [22]. Using this technique, the parameters of the pillar structure are calculated to be r=0.5, $\varepsilon_a$=3.4 and $\varepsilon_b$=1 for the standard unit-cell schematically shown in Fig. 5. EFCs of this crystal are drawn in Fig. 6, showing good conformity with those of the RSC crystal depicted in Fig. 4.

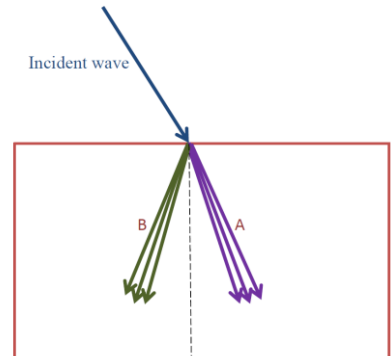

Fig. 3. Schematic representation of a material showing a boundary between positive and negative refraction for the incident spectrum.

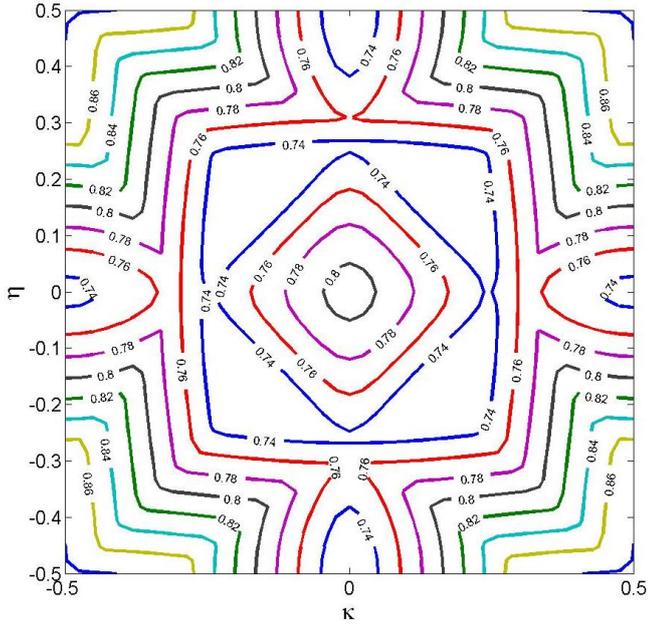

Fig. 4. EFCs of the RSC crystal defined by the parameters in (10). The points specified in (9) are evident in the figure.

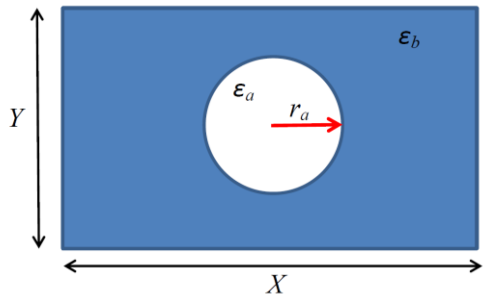

Fig. 5. Unit-Cell of the standard photonic crystal.

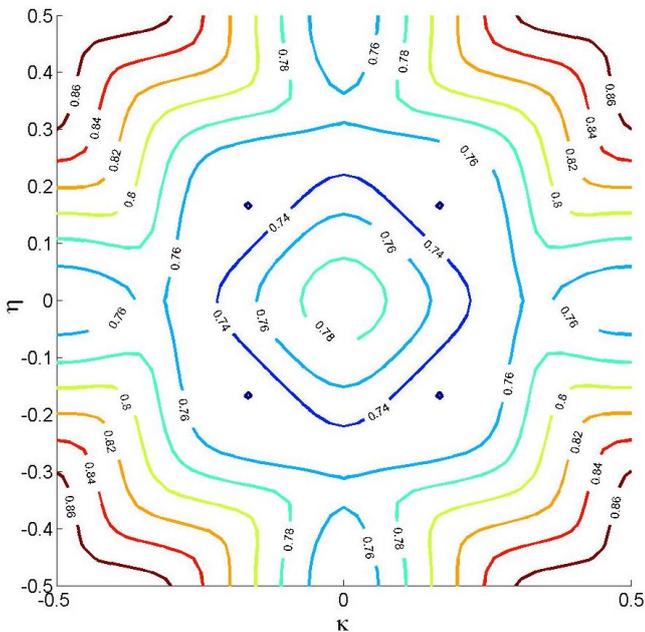

Fig. 6. EFCs of the standard crystal. The EFCs are similar to those of the designed RSC crystal drawn in Fig. 4.

By reducing the unit-cell dimensions from unity to L=0.14μm and calculating the refraction angles of the incident spectrum according to principles outlined in [12], we reach the demultiplexer shown in Fig. 7 with $\lambda_1$=1398nm, $\lambda_2$=1436nm, $\lambda_3$=1474nm, $\lambda_4$=1512nm, and $\lambda_5$=1550nm. As seen in the figure, the designed demultiplexer produces 80-degree beam divergence for 152nm deviation in incident wavelength showing excellent refractive performance.

The aforementioned steps can be repeated to reach other demultiplexer designs. As an example, by choosing $(\kappa_1, \eta_1) = (0,0.1)$ for ω=0.35 and $(\kappa_1, \eta_1) = (0,0.2)$ for ω=0.34 and $(\kappa_1, \eta_1) = (0,0.3)$ for ω=0.36 and following the same steps, the parameters of the standard crystal are calculated to be r=0.47, $\varepsilon_a$=20 and $\varepsilon_b$=5 with the EFCs depicted in Fig. 8. By downscaling unit-cell length to L=527nm, the designed demultiplexer would be able to produce 84-degree beam divergence for wavelength range λ= [1442nm, 1550nm]. Figure 9 shows light trajectories inside the crystal for $\lambda_1$=1442nm, $\lambda_2$=1463nm, $\lambda_3$=1505nm, and $\lambda_4$=1550nm.

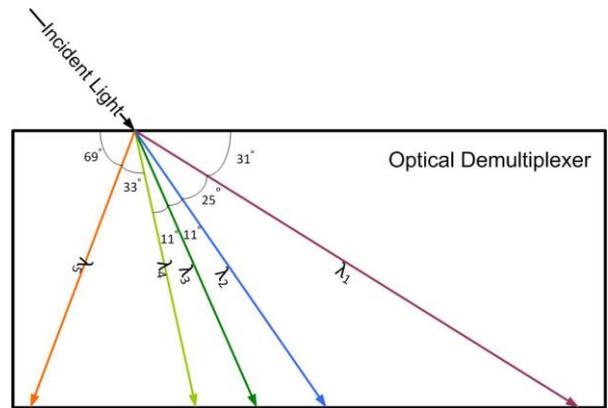

Fig. 7. Refractive properties of the designed demultiplexer in the wavelength range of interest. $\lambda_1$=1398nm, $\lambda_2$=1436nm, $\lambda_3$=1474nm, $\lambda_4$=1512nm and $\lambda_5$=1550nm.

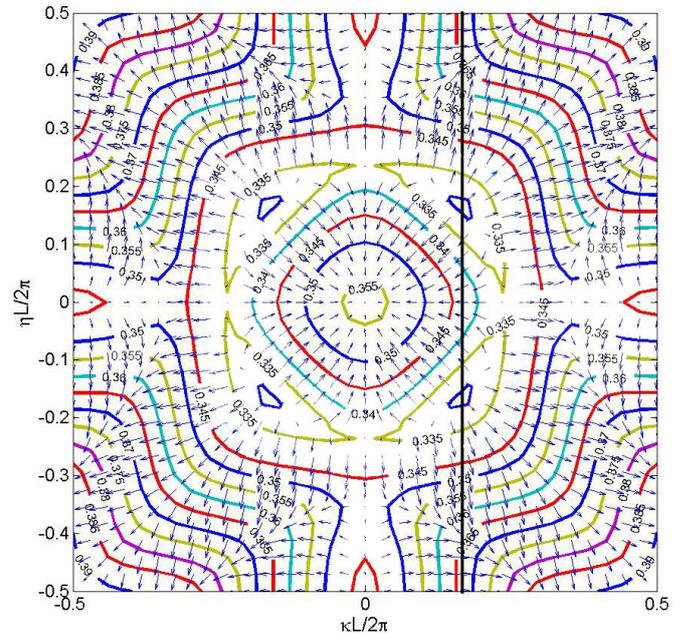

Fig. 8. EFCs of the standard crystal by choosing $(\kappa_1, \eta_1) = (0,0.1)$ for ω=0.35 and $(\kappa_1, \eta_1) = (0,0.2)$ for ω=0.34 and $(\kappa_1, \eta_1) = (0,0.3)$ for ω=0.36.

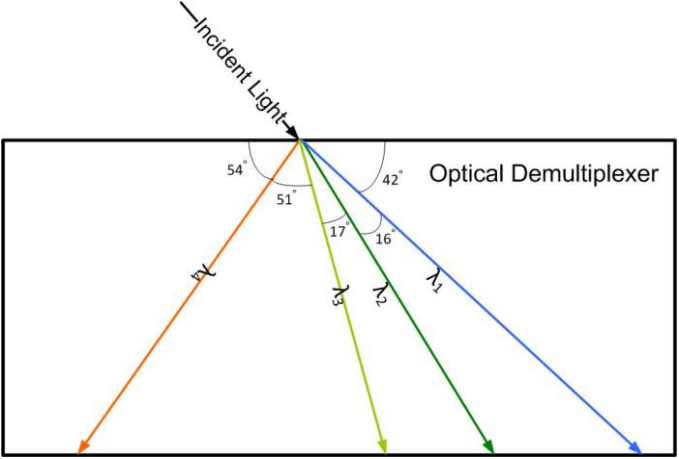

Fig. 9. Refractive properties of the designed demultiplexer in the wavelength range of interest. $\lambda_1$=1442nm, $\lambda_2$=1463nm, $\lambda_3$=1505nm and $\lambda_4$=1550nm.

## IV. TWO-STEP DEMULTIPLEXING

The design technique described in the previous section can be combined with the technique reported in [15] to even further improve the demultiplexing properties of the design. As is clarified in [15], an oblique boundary between two photonic crystals can change the Bloch wavenumber of the propagating wave and change its traveling direction as shown in Fig. 10. The Bloch wavevector in the first crystal can be determined by drawing a vertical line from the initial wavevector. While determining the Bloch wavevector of the second crystal, we should have in mind that the projections of the wavevectors of the two media on the boundary should be equal. Therefore, we should draw a line perpendicular to the slope of the boundary as is shown in Fig 10-b. Schematic EFC of the second crystal together with the Bloch wavevector in the first crystal ($\kappa_1$) and also a line showing the slope of the boundary ($\alpha$) are drawn in Fig. 10-b. A dotted line perpendicular to the boundary is drawn from the end-point of $\kappa_1$. The intersection of this line with the EFC determines the Bloch wavevector in the second crystal ($\kappa_2$). The slope of the boundary can be engineered to excite the desired wavevectors in the second crystal in order to produce a high angular separation between different wavelengths.

An example of employing this technique in demultiplexer design is described in figures 11 and 12. Figure 11 depicts the EFCs of the second photonic crystal. Points 1, 2 and 3 in this figure are the excited Bloch wavevectors in the first photonic crystal. As is shown in Fig. 11, a line perpendicular to the respective boundary (See Fig. 12) is drawn from the Bloch wavevector of each propagating wavelength. The intersections of these lines with the corresponding EFC determine the Bloch wavevectors and therefore the propagation directions in the second crystal. As is seen in Fig. 12, the two-step demultiplexer employing positive-negative refraction boundary in the first crystal and oblique boundaries with the second crystal, produces super demultiplexing behavior, separating a spectrum of $\lambda$= [1474nm, 1550nm] by 161 degrees.

## V. CONCLUSION

In this paper, a systematic design algorithm was employed to create a positive-negative refraction boundary in a photonic crystal to boost the refractive properties of the optical demultiplexer. EFC engineering could be done by a smart selection of three points to be included in the EFC patterns of an RSC structure to guarantee the positive-negative refraction boundary. The RSC crystal was then transformed into a standard pillar structure with the help of mathematical methods based on Fourier analysis. The designed demultiplexer was able to produce 84-degree wavelength separation for the incident spectrum $\lambda$= [1442nm, 1550nm].

The demultiplexer's operational characteristics were even further improved by applying a two-step demultiplexing technique, in which the Bloch wavenumbers of the propagating waves in the first crystal could be changed after incidence on an oblique boundary of a second crystal. Using this technique, a wavelength separation of 163 degrees was obtained for the input wavelength range $\lambda$= [1474nm, 1550nm].

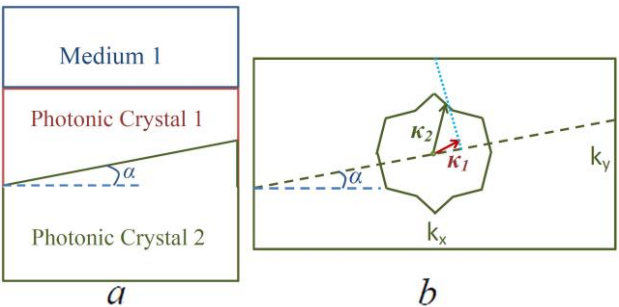

Fig. 10. (a) A hetero photonic crystal structure with oblique boundaries. (b) The excited wavevector inside the second crystal ($\kappa_2$) can be determined by the intersection of a line drawn from the endpoint of the wavevector of the first crystal ($\kappa_1$) perpendicular to the slope of the boundary with the crystal EFC.

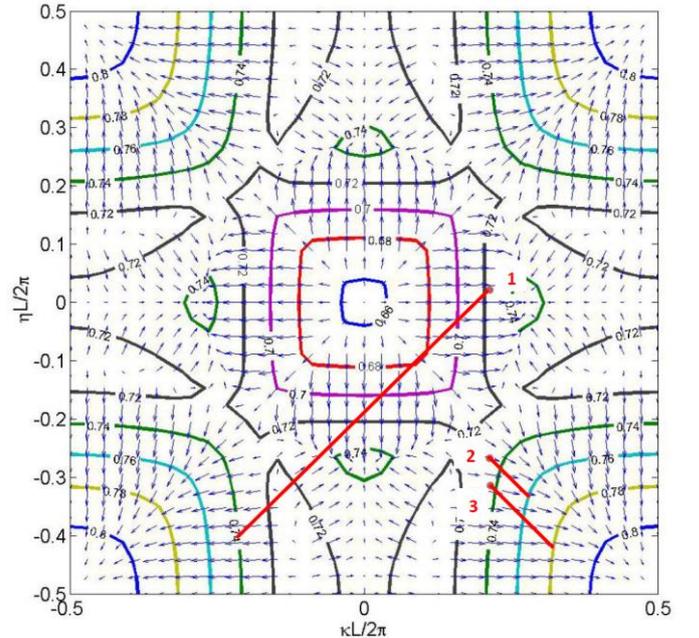

Fig. 11. EFCs of the second crystal in the two-step demultiplexer design shown in Fig. 12. Points 1,2 and 3 are the excited Bloch wavevectors in the first photonic crystal. The intersection of the perpendicular line to the corresponding boundary with the corresponding EFC determines the excited Bloch wavevector in the second photonic crystal.

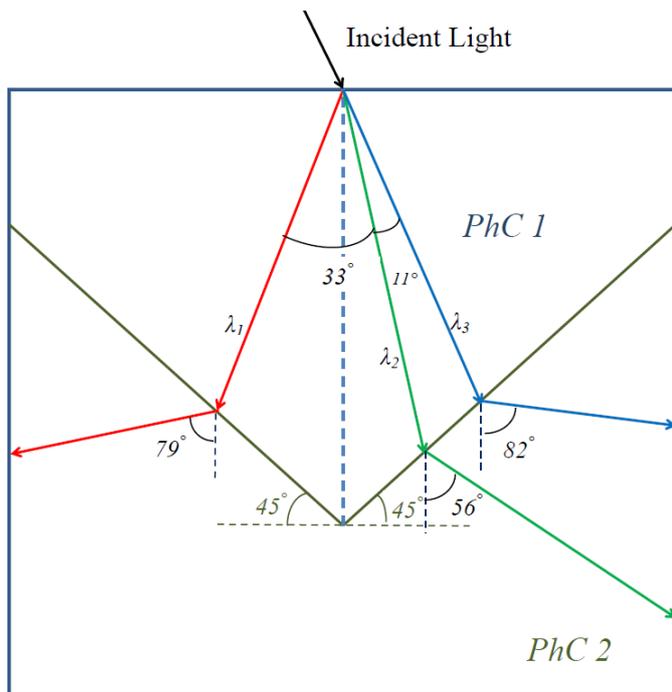

Fig. 12. Refractive properties of the two-step demultiplexer in the wavelength range of λ=[1474nm, 1550nm]. The first photonic crystal exhibits a positive-negative refraction boundary and separates the incident spectrum by 44 degrees. The oblique boundary between the two crystals leads to a change of Bloch wavevectors in the second crystal, thereby improving the wavelength separation to 161 degrees.
## References

[1] S. Lin, V. Hietala, L. Wang and E. Jones, "Highly dispersive photonic band-gap prism," *Optics Letters,* vol. 21, no. 21, pp. 1771-1773, November 1996.

[2] H. Kosaka, T. Kawashima, A. Tomita, M. Notomi, T. Tamamura, T. Sato and S. Kawakami, "Superprism phenomena in photonic crystals," *Physical Review B,* vol. 58, no. 16, pp. R10096-R10099, October 1998.

[3] H. Kosaka, T. Kawashima, A. Tomita, M. Notomi, T. Tamamura, T. Sato and S. Kawakami, "Superprism Phenomena in Photonic Crystals: Toward Microscale Lightwave Circuits," *Jounal of Lightwave Technology,* vol. 17, no. 11, pp. 2032-2038, November 1999.

[4] L. Wu, M. Mazilu and T. F. Krauss, "Beam steering in planar-photonic crystals: from superprism to supercollimator," *Journal of Lightwave Technology,* vol. 21, no. 2, pp. 561-566, February 2003.

[5] T. Baba and M. Nakamura, "Photonic Crystal Light Deflection Devices Using the Superprism Effect," *IEEE Journal of Quantum Electronics,* vol. 38, no. 7, pp. 909-914, July 2002.

[6] P. Halevi, A. A. Krokhin and J. Arriaga, "Photonic crystal optics and homogenization of 2D periodic composites," *Physical Review Letters,* vol. 82, no. 4, pp. 719-722, January 1999.

[7] S. H. Kim, H. S. Park, J. H. Choi, J. W. Shim and S. M. Yang, "Integration of colloidal photonic crystals toward miniaturized spectrometers," *Advanced Materials,* vol. 22, no. 9, pp. 946-950, March 2010.

[8] X. Lin, X. Zhang, L. Chen, M. Soljacic and X. Jiang, "Super-collimation with high frequency sensitivity in 2D photonic crystals induced by saddle-type van Hove singularities," *Optics Express,* vol. 21, no. 5, pp. 30140-30147, December 2013.

[9] M. Turduev, I. H. Giden and H. Kurt, "Extraordinary wavelength dependence of self-collimation effect in photonic crystal with low structural symmetry," *Photonics and Nanostructures – Fundamentals and Applications,* vol. 11, no. 3, p. 241–252, August 2013.

[10] W. Li, X. Zhang, X. Lin and X. Jiang, "Enhanced wavelength sensitivity of the self-collimation superprism effect in photonic crystals via slow light," *Optics Express,* vol. 39, no. 15, pp. 4486-4489, August 2014.

[11] B. Smida and P. Pahlavan, "Unified Wireless Power and Information Transfer Using a Diplexed Rectifier," in *IEEE Global Communications Conference (GLOBECOM)*, Madrid, 2021.

[12] M. Notomi, "Theory of light propagation in strongly modulated photonic crystals: Refraction like behavior in the vicinity of the photonic band gap," *Physical Review B,* vol. 62, no. 16, p. R10696–R10705, October 2000.

[13] B. Gralak, S. Enoch and G. Tayeb, "Anomalous refractive properties of photonic crystals," *Journal of Optical Soceity of America A,* vol. 17, no. 6, p. 1012–1020, June 2000.

[14] S. Pahlavan and V. Ahmadi, "A Systematic Approach to Photonic Crystal Based Metamaterial Design," *International Journal of Optics and Photonics,* vol. 10, no. 1, pp. 55-64, 2016.

[15] S. Pahlavan and V. Ahmadi, "Novel Optical Demultiplexer Design Using Oblique Boundary in Hetero Photonic Crystals," *Photonic Technology Letters,* vol. 29, no. 6, pp. 511-514, March 2017.

[16] P. Pahlavan, S. Z. Aslam and N. Ebrahimi, "A Novel Dual-band and Bidirectional Nonlinear RFID Transponder Circuitry," in *2022 IEEE/MTT-S International Microwave Symposium*, Denver, 2022.

[17] N. A. Nicorovici, R. C. McPhedran and L. C. Botten, "Photonic band gaps: Noncommuting limits and the Acoustic Band," *Physical Review Letters,* vol. 75, no. 8, pp. 1507-1510, 1995.

[18] R. C. McPhedran, N. A. Nicorovici and L. C. Botten, "The TEM mode and homogenization of doubly periodic structures," *Journal of Electromagnetic Waves and Applications,* vol. 11, no. 7, pp. 981-1012, 1997.

[19] A. Yariv and P. Yeh, Optical Waves in Crystals, New York: Wiley, 1984.

[20] J. P. Dowling and C. M. Bowden, "Anomalous index of refraction in photonic bandgap materials," *Journal of Modern Optics ,* vol. 41, no. 2, pp. 345-351, 1994.

[21] S. Enoch, G. Tayeb and D. Maystre, "Numerical evidence of ultrarefractive optics in photonic crystals," *Optics Communications*, vol. 161, no. 4-6, pp. 171-176, 1999.

[22] S. Khorasani and A. Adibi, "Approximate analysis and design of rectangular-lattice photonic crystals," *Optics Letters,* vol. 28, no. 16, pp. 1472-1474, 2003.

[23] M. J. Powell, A Fortran Subroutine for Solving Systems of Nonlinear Algebraic Equations, United Kingdom: https://www.osti.gov/servlets/purl/4772677, 1968.

[24] K. Sakoda, Optical Properties of Photonic Crystals, Springer, 2004.

[25] P. R. Villeneuve and M. Piché, "Photonic band gaps in two-dimensional square and hexagonal lattices," *Physical Review B,* vol. 46, no. 8, pp. 4969-4972, 1992.